\documentclass{desyproc}
\newcommand{\rr}{\mbox{\boldmath $r$}}

\begin{document}
\title{DVCS at HERA and at CERN}

\author{{\slshape Laurent Schoeffel}\\[1ex]
CEA Saclay, Irfu/SPP, 91191 Gif-sur-Yvette Cedex, France }

\contribID{48}

\confID{1407}  
\desyproc{DESY-PROC-2009-03}
\acronym{PHOTON09} 
\doi  

\maketitle

\begin{abstract}
Deeply Virtual Compton Scattering (DVCS)
in $ep$ collisions has emerged in recent years as a an
essential reaction to obtain information on the correlation
of partons in the hadron (proton) or on the transverse distribution
of these partons.
In these proceedings, we examine the lattest data from HERA 
(at low $x_{Bj}<10^{-2}$) and their
impact on models. We analyse in detail what these data
imply on the spatial structure of the proton.
In particular,
the most recent measurements of the Beam Charge Asymmetry 
by the H1 experiment is discussed in this context.
Perspectives are presented 
for further measurements of
DVCS cross sections at CERN, within the COMPASS experiment.
\end{abstract}

\section{Introduction}

Measurements of the deep-inelastic scattering (DIS) of leptons and nucleons, $e+p\to e+X$,
allow the extraction of Parton Distribution Functions (PDFs) which describe
the longitudinal momentum carried by the quarks, anti-quarks and gluons that
make up the fast-moving nucleons. 
While PDFs provide crucial input to
perturbative Quantum Chromodynamic (QCD) calculations of processes involving
hadrons, they do not provide a complete picture of the partonic structure of
nucleons. 
In particular, PDFs contain neither information on the
correlations between partons nor on their transverse motion.

Hard exclusive processes, in  which the
nucleon remains intact, have emerged in recent years as prime candidates to complement
this essentially one dimentional picture. 
The simplest exclusive process is the deeply virtual
Compton scattering (DVCS) or exclusive production of real photon, 
$e + p \rightarrow e + \gamma + p$.
This process is of particular interest as it has both a clear
experimental signature and is calculable in perturbative QCD. 
The DVCS reaction can be regarded as the elastic scattering of the
virtual photon off the proton via a colourless exchange, producing a 
real photon in the final state  \cite{dvcsh1,dvcszeus}. 
In the Bjorken scaling 
regime, 
QCD calculations assume that the exchange involves two partons, having
different longitudinal and transverse momenta, in a colourless
configuration. These unequal momenta or skewing are a consequence of the mass
difference between the incoming virtual photon and the outgoing real
photon. This skewedness effect can
 be interpreted in the context of generalised
parton distributions (GPDs) \cite{qcd} or dipole model \cite{dipole}. 
In the following, we examine the most recent data recorded from the DESY $ep$
collider at HERA and their implication
on the quarks/gluons imaging of the nucleon \cite{dvcsh1,dvcszeus}.

\section{Latest experimental measurements from HERA}

\begin{figure}[!] 
  \begin{center}
    \includegraphics[width=6cm]{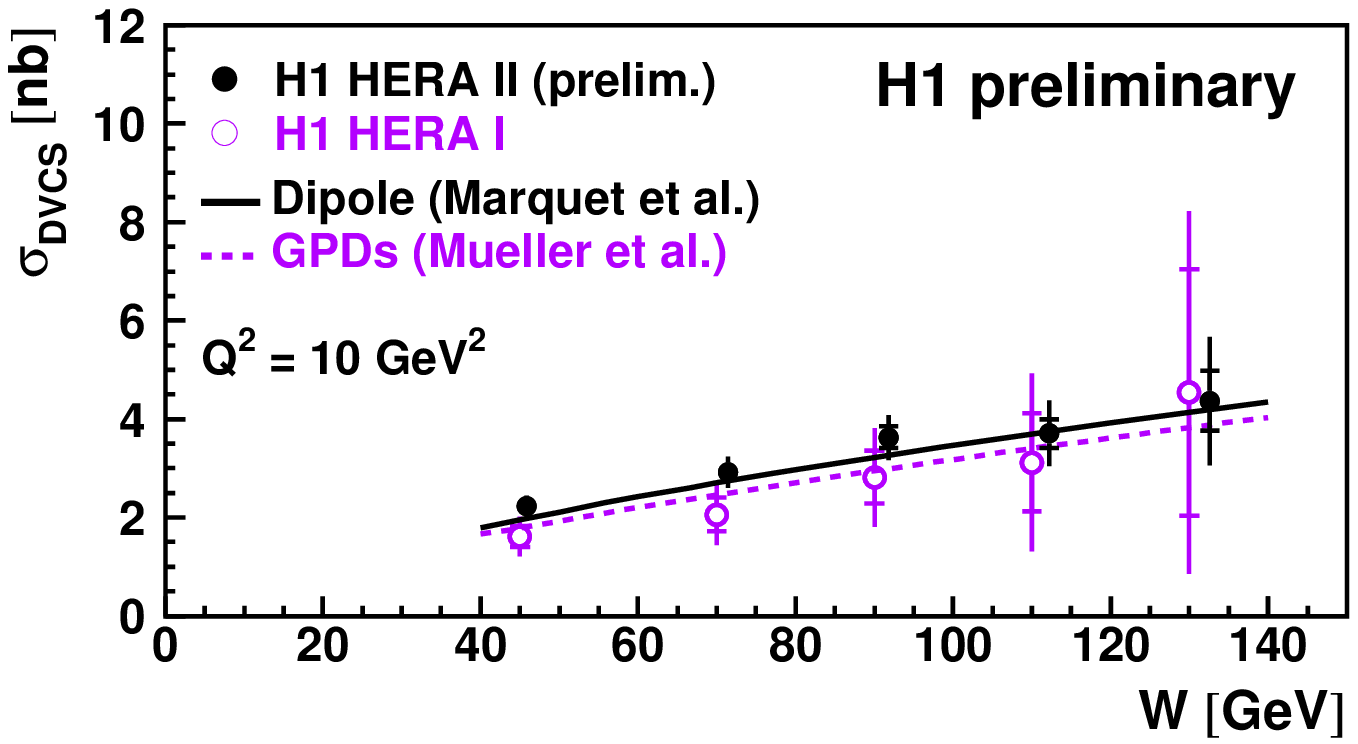}
    \includegraphics[width=6cm]{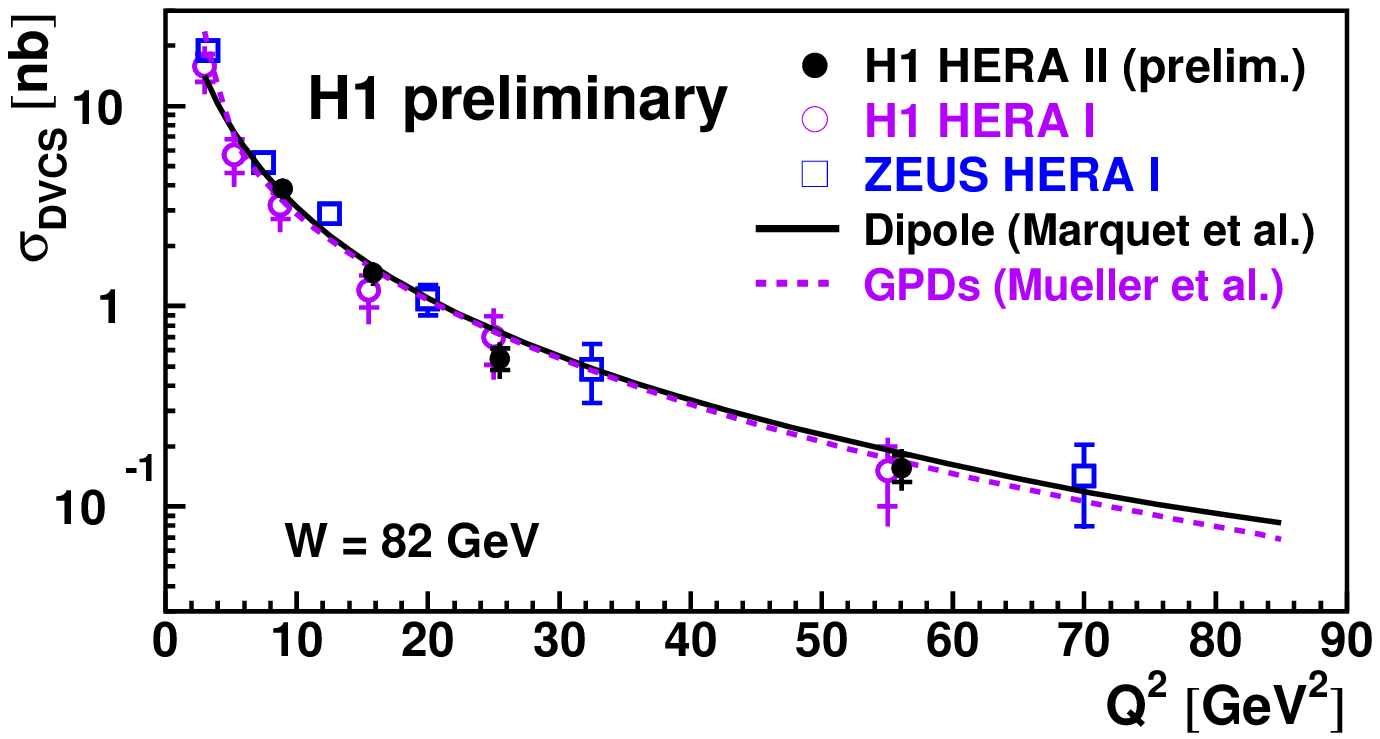}
  \end{center}
  \caption{DVCS cross section for the full HERA data as a function of
$W$ and $Q^2$.
}
\label{fig1}  
\end{figure}

The first measurements of DVCS cross section have been realised  at HERA within 
the H1 and
ZEUS experiments \cite{dvcsh1,dvcszeus}. 
These results are given in the specific
 kinematic domain
of both experiments,
at low $x_{Bj}$ ($x_{Bj} < 0.01$) but they take advantage of the large range in $Q^2$, 
offered by the
HERA kinematics, which covers more than 2 orders
of magnitude, from $1$ to $100$ GeV$^2$. It makes possible to study the transition from
the low $Q^2$ non-perturbative region (around $1$ GeV$^2$) towards higher values 
of $Q^2$ where the higher twists
effects are lowered (above $10$ GeV$^2$).
The last DVCS cross sections as a functon of
$Q^2$ and $W \simeq \sqrt{Q^2/x}$ are presented 
on Fig. \ref{fig1}. A good agreement with GPDs \cite{qcd} 
and dipole \cite{dipole} models is observed.
A very fundamental observation is the steep $W$ dependence in $W^{0.7}$, visible
on Fig. \ref{fig1}. This means that DVCS is a hard process. Thus, 
it is justified to compare DVCS measurements with perturbative QCD calculations,
 GPDs or dipole approaches, as displayed in Fig. \ref{fig1}.

\section{The colour dipole model}

Let's discuss in more details the colour dipole model.
Indeed, this approach  provides a simple 
unified picture of inclusive and diffractive processes and enables hard
and soft physics to be incorporated in a single dynamical framework. 
At high energies, in the proton's rest frame, the virtual photon fluctuates 
into a hadronic system (the simplest of which is a $q {\bar q}$ dipole) a 
long distance upstream of the target proton. The formation time of this hadronic 
system, and of the subsequent formation of the hadronic final state, is much longer 
than the interaction time with the target. 

DVCS is a good probe of the transition between soft and hard regimes in 
the dipole model for two reasons. Indeed, the transverse photon 
wave function can select large dipoles, even for large $Q^{2}$, and certainly 
for the $Q^2$ range $2 < Q^2 < 20$ GeV$^2$. Also, because the final photon is real,
DVCS is more sensitive to large dipoles than DIS at the same $Q^2$.
Then, in the colour dipole approach, the DIS (or DVCS) process can be seen as a
succession in time of three factorisable subprocesses: i)  the virtual
photon fluctuates in a quark-antiquark pair, ii) this colour dipole
interacts with the proton target, iii) the quark pair annihilates into a
virtual (or real) photon. The imaginary part of the DIS (or DVCS)
amplitude at $t=0$ is expressed in the simple way \cite{dipole,marquet}
\begin{eqnarray} 
 {\cal I}m\, {\cal A}\,(W,Q_1,Q_2)  =  \sum_{T,L}\int \limits_0^1 dz \! 
 \int_{0}^{\infty} d^2\rr\, 
 \Psi_{T,L}^*(z,\,\rr,\,Q_1^2)\,\sigma_{dip}\,(z,\rr)\Psi_{T,L}
 (z,\,\rr,\,Q_2^2)\label{dvcsdip}\,,
\end{eqnarray} 
where $\Psi(z,\rr,Q_{1,2})$ are the light cone photon wave functions for
transverse and longitudinal photons. The quantity $Q_1$ is the virtuality
of the incoming photon, whereas $Q_2$ is the virtuality of the outgoing
photon. 
In the DIS case, one has $Q_1^2=Q_2^2=Q^2$ and for DVCS,
$Q_1^2=Q^2$ and $Q_2^2=0$. 
The relative transverse quark pair (dipole) separation is labeled by 
$\rr$ whilst $z$ (respec.\ $1-z$) labels the quark (antiquark)
longitudinal momentum fraction.

It should be noticed that the non-forward kinematics for DVCS is encoded in
the colour dipole approach through the different weight coming from the
photon wavefunctions in Eq. (\ref{dvcsdip}). The off-diagonal effects,
which affect the gluon and quark distributions in GPDs models,
should be included in the parameterisation of the dipole cross section. At
the present stage of the development of the  dipole formalism, we
have no accurate theoretical arguments on how to compute skewedness effects
from first principles. A consistent approach would be to compute the
scattering amplitude in the non-forward case, since the non-forward photon
wave function has been recently obtained.
In this case, the dipole cross section,
$\sigma_{dip}\,(x_1,x_2,\rr,\vec{\Delta })$, depends on the 
momenta $x_1$ and $x_2$ carried by the exchanged gluons, respectively, and
on the total transverse momentum transfer $\vec{\Delta}$. In this case,
additional information about the dependence upon $\vec{\Delta}$ is needed for
the QCD Pomeron and proton impact factor.  A first attempt in this direction
is done in \cite{dipole}.

\begin{figure}[!htbp]
\begin{center}
 \includegraphics[totalheight=5.8cm]{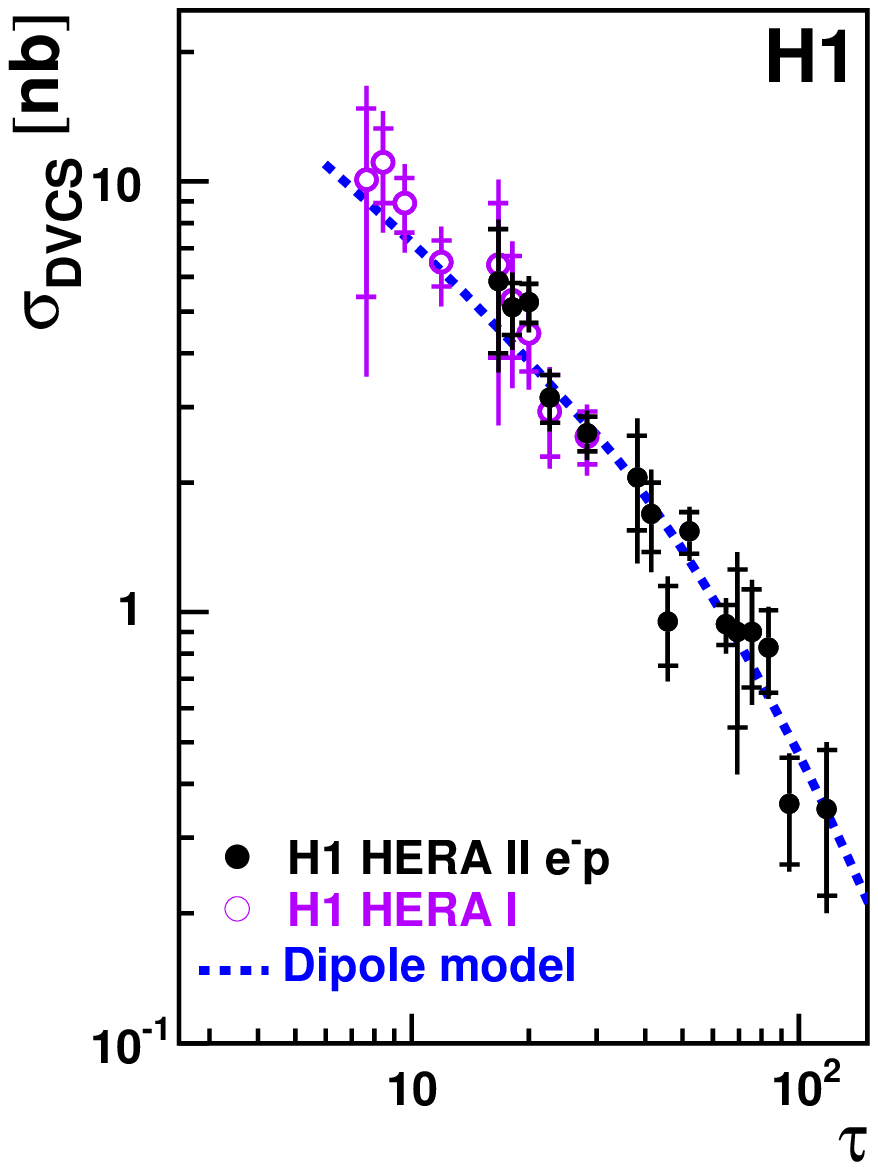}
 \includegraphics[totalheight=5.8cm]{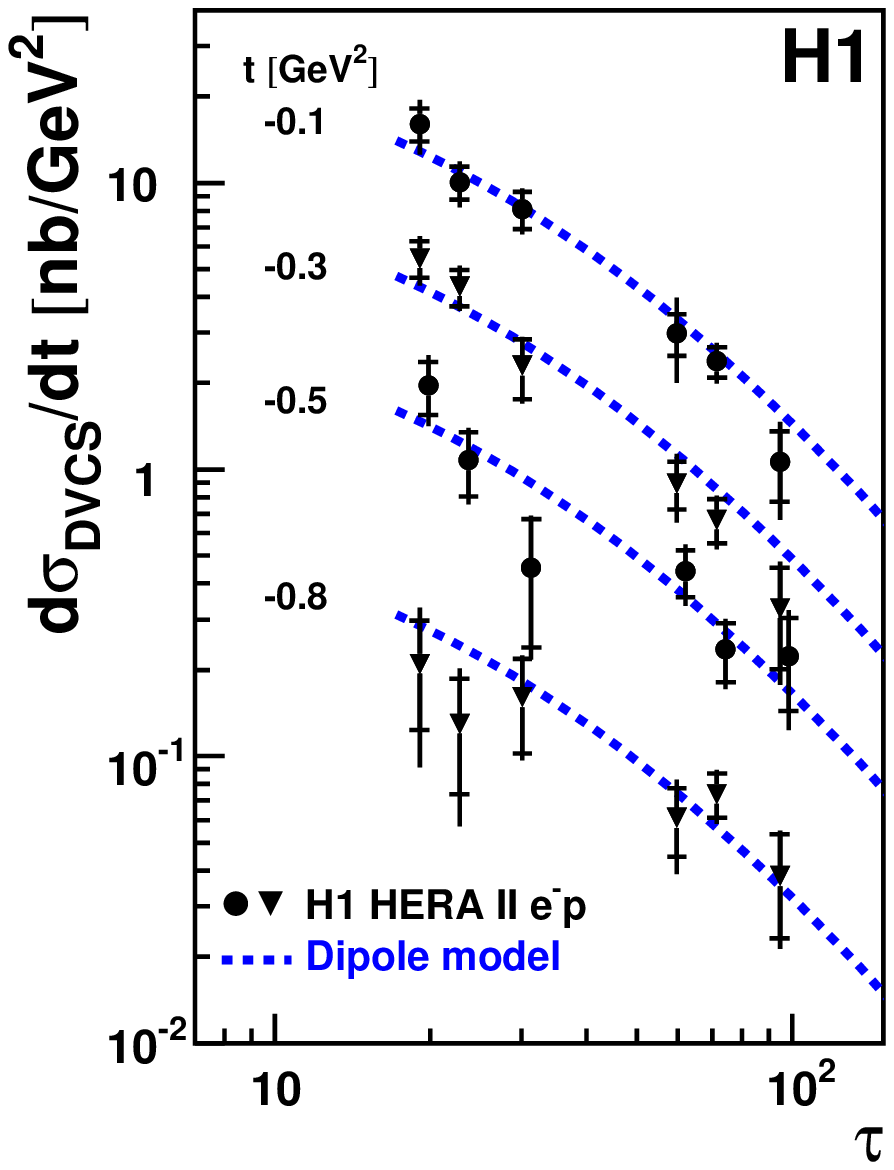}
\end{center}
\caption{\label{figdipole} 
DVCS cross section measurements as a function of 
the scaling variable
$\tau={Q^2}/{Q_s^2(x)}$. 
Results are shown
for the full $t$ range  $|t| <$ 1 GeV$^2$ (left)
and at four values of $t$ (right).
The dashed curves represent the predictions of the 
dipole model \cite{marquet}.
}
\end{figure}

At very small values of the Bjorken scaling variable $x$
the saturation regime of QCD can be reached.
In this domain, the gluon density in the proton is so large that 
non-linear effects like gluon recombination tame its growth.
In the dipole model approach, the transition to the saturation regime is characterised by the so-called 
saturation scale parametrised here as $Q_s(x)=Q_0 ({{x_0}/{x}})^{-\lambda/2}$, where $Q_0$, $x_0$ and $\lambda$ are parameters. 
The transition to saturation occurs when $Q$ becomes comparable to 
$Q_s(x)$.
An important feature of dipole models that incorporate saturation is that the total cross section can be expressed as a function of the single variable $\tau$:

\begin{equation}
\sigma_{tot}^{\gamma^\ast p}(x,Q^2)  = \sigma_{tot}^{\gamma^\ast p}( \tau ) , \;\;  \mbox{with} \; \ \ \ \tau=\frac{Q^2}{Q_s^2(x)}.
\label{eq:tau}
\end{equation}
This property, called geometric scaling, has already been observed  to hold
for the total $ep$ DIS cross section  and in diffractive processes \cite{marquet} (see Fig. \ref{figdipole}). 
It has also recently been addressed in the context of
exclusive processes including DVCS and extended to cases with non-zero momentum transfer to the proton \cite{dipole}.
It is therefore interesting to test if the present DVCS measurements obey the geometric scaling laws predicted by such models,
as illustraded in Fig. \ref{figdipole} (right plot for non-zero momentum transfer to the proton).

\section{Nucleon Tomography and Perspectives at CERN}

\begin{figure}[htbp] 
  \begin{center}    
    \includegraphics[width=6cm]{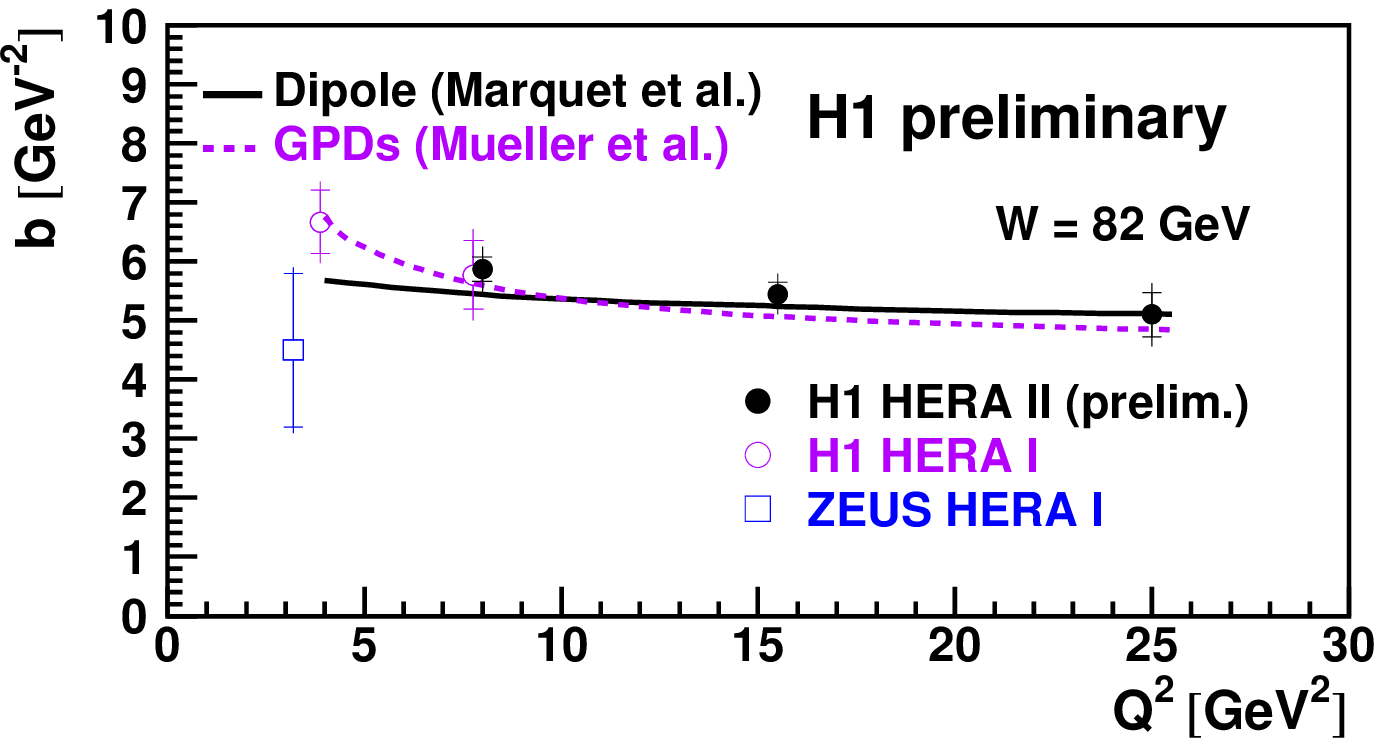}
    \includegraphics[width=6cm]{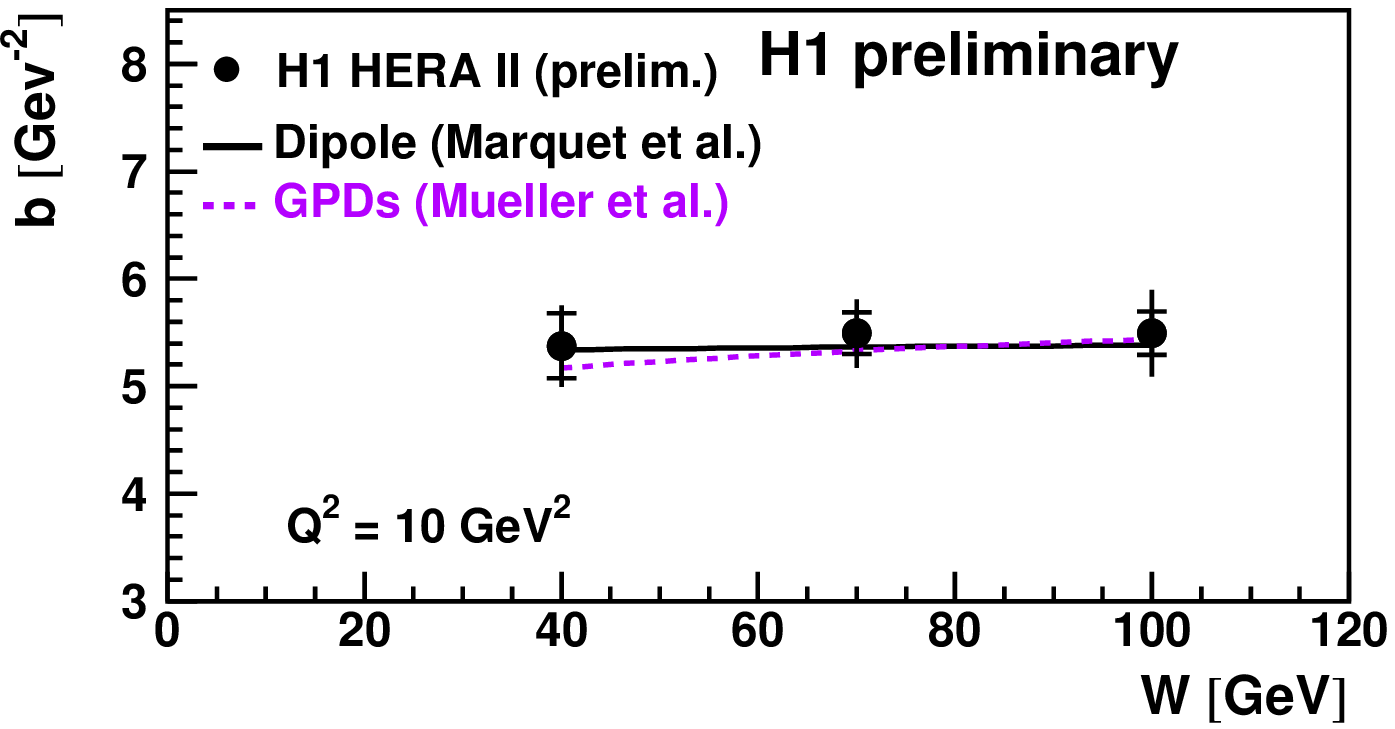}
  \end{center}
  \caption{The logarithmic slope of the $t$ dependence
  for DVCS exclusive production, $b$ as a function of $Q^2$ and $W$, extracted from a fit
  $d\sigma/dt \propto
\exp(-b|t|)$  where $t=(p-p')^2$.
}
\label{fig2}  
\end{figure}

A major experimental achievement 
of H1 and ZEUS \cite{dvcsh1,dvcszeus} has been the measurement of
DVCS cross sections, differential in $t=(p'-p)^2$, 
the momentum transfer (squared) at the proton vertex.
A good description
of $d\sigma_{DVCS}/dt$ by a fit of the form $e^{-b|t|}$
is obtained \cite{dvcsh1,dvcszeus}. 
Hence, an extraction of the $t$-slope parameter $b$ is accessible
and it can be achieved experimentally
for different values of $Q^2$ and $W$ (see Fig. \ref{fig2}).
Again, we observe the good agreement 
of measurements with GPDs and dipole models.

\begin{figure}[!htbp] 
  \begin{center}
    \includegraphics[width=8cm]{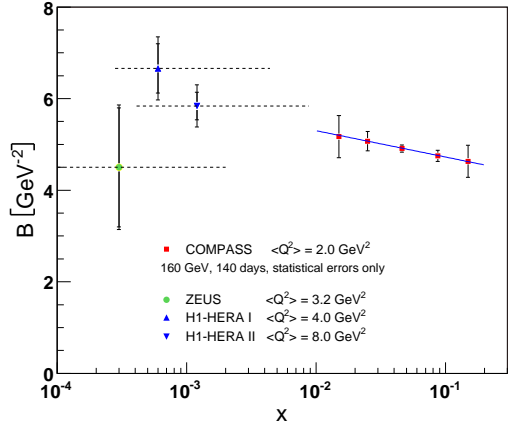}
  \end{center}
  \caption{The logarithmic slope of the $t$ dependence
  for DVCS exclusive production, $b$ as a function of $x_{Bj}$, 
  extracted from a fit
  $d\sigma/dt \propto \exp(-b|t|)$  where $t=(p-p')^2$.
  H1 and ZEUS points are measurements and simulations are
  displayed for COMPASS (CERN).
}
\label{fig2b}  
\end{figure} 

Measurements of the $t$-slope parameters $b$
are key measurements for almost all exclusive processes,
in particular DVCS.
 Indeed,
a Fourier transform from momentum
to impact parameter space readily shows that the $t$-slope $b$ is related to the
typical transverse distance between the colliding objects \cite{qcd}.
At high scale, the $q\bar{q}$ dipole is almost
point-like, and the $t$ dependence of the cross section is given by the transverse extension 
of the gluons (or sea quarks) in the  proton for a given $x_{Bj}$ range.
More precisely, from GPDs, we can compute
a parton density which also depends on a spatial degree of freedom, the transverse size (or impact parameter), labeled $R_\perp$,
in the proton. Both functions are related by a Fourier transform 
$$
PDF (x, R_\perp; Q^2) 
\;\; \equiv \;\; \int \frac{d^2 \Delta_\perp}{(2 \pi)^2}
\; e^{i ({\Delta}_\perp {R_\perp})}
\; GPD (x, t = -{\Delta}_\perp^2; Q^2).
$$
Thus, the transverse extension $\langle r_T^2 \rangle$
 of gluons (or sea quarks) in the proton can be written as
$$
\langle r_T^2 \rangle
\;\; \equiv \;\; \frac{\int d^2 R_\perp \; PDF(x, R_\perp) \; R_\perp^2}
{\int d^2 R_\perp \; PDF(x, R_\perp)} 
\;\; = \;\; 4 \; \frac{\partial}{\partial t}
\left[ \frac{GPD (x, t)}{GPD (x, 0)} \right]_{t = 0} = 2 b
$$
where $b$ is the exponential $t$-slope.
Measurements of  $b$
presented in Fig. \ref{fig2}
corresponds to $\sqrt{r_T^2} = 0.65 \pm 0.02$~fm at large scale 
$Q^2$ for $x_{Bj} < 10^{-2}$.
This value is smaller that the size of a single proton, and, 
in contrast to hadron-hadron scattering, it does not expand as energy $W$ increases.
This result is consistent with perturbative QCD calculations in terms 
of a radiation cloud of gluons and quarks
emitted around the incoming virtual photon.
The fact the perturbative QCD calculations provide correct descriptions
of $b$ measurements (see previous section) is a proof that
they deal correctly this this non-trivial aspect of the proton 
(spatial) structure. 
As already discussed in the context of dipole models,
the modeling of the correlation between the spatial transverse
structure and the longitudinal momenta distributions of partons in the proton
is one major challenge also for the GPDs approach. 
In Fig. \ref{fig2b}, we present a summary of the measurements of 
$b(x_{Bj})$ by
H1 and ZEUS experiments and simulations for COMPASS at CERN.
The importance of such measurements of $t$-slopes as a function
of $x_{Bj}$ at CERN is then obvious.

\begin{figure}[!htbp] 
  \begin{center}
    \includegraphics[width=6cm]{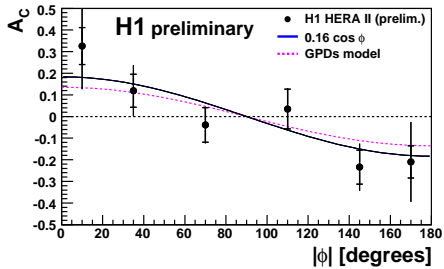}
  \end{center}
  \caption{Beam charge asymmetry as a function of $\phi$ measured by H1.
  Statistical and systematical uncertainties are shown. Data are corrected 
  from the migrations of events in $\phi$.
  A comparison with the GPDs model described in Ref. \cite{qcd} is presented.
  It fits very nicely with the best fit to the data, in $p_1 \cos \phi$ ($p_1=0.16$).
}
\label{fig3}  
\end{figure}

Another natural  way to address the problem 
of the correlation  between $x$ and $t$ kinematical variables
proceeds from
a determination of a cross section asymmetry with respect to the beam
charge. It has been realised recently by the H1 experiment by measuring the ratio
$(d\sigma^+ -d\sigma^-)/ (d\sigma^+ + d\sigma^-)$ as a function of $\phi$,
where $\phi$ is the azimuthal angle between leptons and proton plane.
The result is presented on Fig. \ref{fig3} with  a fit in $\cos \phi$.
After applying a deconvolution method to account for the  resolution on $\phi$,
the coefficient of the $\cos \phi$ dependence is found to 
be $p_1 = 0.16 \pm 0.03 (stat.) \pm 0.05 (sys.)$ (at low $x_{Bj}<0.01$).
This result represents  a major experimental progress 
and is challenging for models (see Fig. \ref{fig3}). 
Let's note that models of GPDs can use
 present HERA data at low $x_{Bj}$,
as well as JLab and HERMES data at larger $x_{Bj}$ ($x_{Bj}>0.1$),
in order to provide a first global understanding of exclusive real photon
production \cite{qcd}.  
However, as already mentioned above, some efforts have
still to be made in the intermediate $x_{Bj}$ domain.

\begin{figure}[htbpp]
{\centering \resizebox*{0.4\textwidth}{!}{\includegraphics{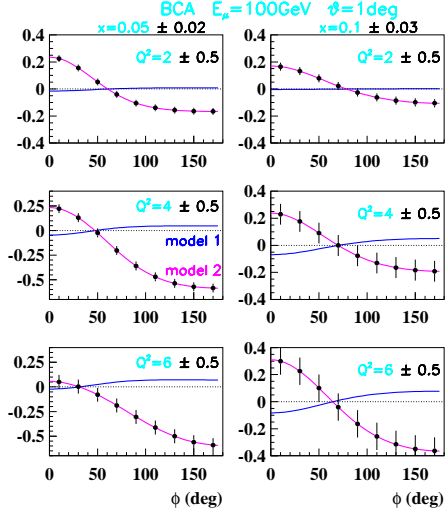}} \par}
\caption{\label{compass2}  Azimuthal distribution of the beam charge asymmetry
measured at COMPASS at \protect\( E_{\mu }\protect \)= 100 GeV and
\protect\( |t|\leq 0.6\protect \) GeV\protect\( ^{2}\protect \)
for 2 domains of \protect\( x_{Bj}\protect \) (\protect\( x_{Bj}=0.05\pm 0.02\protect \)
and \protect\( x_{Bj}=0.10\pm 0.03\protect \)) and 3 domains of \protect\( Q^{2}\protect \)
(\protect\( Q^{2}=2\pm 0.5\protect \) GeV\protect\( ^{2}\protect \),
\protect\( Q^{2}=4\pm 0.5\protect \) GeV\protect\( ^{2}\protect \)
and \protect\( Q^{2}=6\pm 0.5\protect \) GeV\protect\( ^{2}\protect \))
obtained in 6 months of data taking with a global efficiency of 25\%
and with \protect\( 2\cdot 10^{8}\protect \) \protect\( \mu \protect \)
per SPS spill (\protect\( P_{\mu ^{+}}=-0.8\protect \) and \protect\( P_{\mu ^{-}}=+0.8\protect \)).
More recent simulations are on going which does not change the conclusions of this plot. }
\end{figure}

Feasabilities for future Beam Charge Asymmetry (BCA) 
measurements at COMPASS have been studied extensively
in the last decade \cite{dhose}.
 COMPASS is a fixed target experiment which can use
100 GeV muon beams and hydrogen targets, and then access 
experimentaly the DVCS process $\mu p \rightarrow \mu \gamma p$.
The BCA can be determined when using positive and negative muon beams.
One major interest is the kinematic coverage from $2$ GeV$^2$ till $6$ GeV$^2$ in $Q^2$
and  $x_{Bj}$ ranging from $0.05$ till $0.1$. It means that it is possible to avoid
the kinematic domain dominated by higher-twists and non-perturbative effects 
(for $Q^2 < 1$ GeV$^2$) and keeping a
$x_{Bj}$ range which is extending the HERA (H1/ZEUS) domain.
As mentioned above, this is obviously an essential measurement to
cover the full kinematic range and give some 
results in the intermediate $x_{Bj}$ range between H1/ZEUS  and 
JLab/HERMES experiments.
Simulations done for COMPASS \cite{dhose,compasslolo} are shown in Fig.
\ref{compass2} for BCA  in a  setup described in the legend of the figure.
Two models of GPDs, with
a factorised and non-factorised $t$ dependence,
are shown in Fig. \ref{compass2} and we can observe easily the great discrimination
power offered by  COMPASS, with the proton recoil detector fully operational.
Of course, the discrimination is large in  Fig.  \ref{compass2} due to the fact that
$\alpha'$ is taken to be large ($\alpha' \sim 0.8$ GeV$^{-2}$) in
 simulations.
If it happens to be much smaller, as measured at low $x_{Bj}$ by H1 \cite{dvcsh1},
both predictions for BCA in Fig.  \ref{compass2} would be of similar shape,
as both curves would converge to the factorised case.
This shows clearly the
high level of sensitivity of this experimental quantity on the modeling
of GPDs. This makes this observable very interesting and challenging
for GPDs models in the future, once these measurements at CERN
would be realised in 2011/2012.


\section{Summary and outlook}
DVCS measurements in the HERA kinematics at low $x_{Bj}$ ($x_{Bj}<0.01$)
are well described by recent GPDs models, which also describe
correctly measurements at larger values of $x_{Bj}$ in
the JLab kinematics.
DVCS measurements in the HERA kinematics 
are also nicely described
within a dipole approach,
which encodes
the non-forward kinematics for DVCS only 
through the different weights coming from the
photon wavefunctions.  Recently, H1 and ZEUS 
experiments have also shown that proton 
tomography at low $x_{Bj}$
enters into the experimental domain of high energy physics, with a first 
experimental evidence
that gluons are located at the periphery of the proton. A new frontier in 
understanding
this structure would be possible at CERN within the 
COMPASS experimental setup. Major advances have already been done
on the design of the project and simulation outputs.



\begin{footnotesize}



%

\end{footnotesize}


\end{document}